%% file: main.tex
\documentclass[conference]{IEEEtran}
\IEEEoverridecommandlockouts
\usepackage{stmaryrd} 
\usepackage{amscd}
\usepackage[all,cmtip,color]{xy}
\usepackage{url}
\usepackage{comment}
\usepackage{algorithm}
\usepackage{algpseudocode}
\usepackage{amsmath}
\usepackage{multirow}
\usepackage{multicol}
\usepackage{xcolor}
\usepackage{xspace}
\usepackage{csquotes}
\usepackage[table]{xcolor}
\usepackage{tabularx}
\usepackage{nicematrix}
\usepackage{amsthm}
\usepackage{mdframed}
\usepackage{amssymb}
\usepackage{listings}
\usepackage[font=small]{caption}
\usepackage{booktabs, tabularx}
\usepackage[binary-units]{siunitx}
\usepackage{graphicx}
\usepackage{subcaption}
\usepackage{import}
\usepackage{tikz}
\usepackage{hyperref} 
\usepackage{cleveref}
\usepackage{threeparttable}
\usepackage{soul}
\usepackage{enumitem}
\usepackage{soul}
\usepackage{nicematrix} 
\usepackage{amsmath} 

\usepackage{hyperref}
\usepackage{xurl}
\hypersetup{breaklinks=true}

\newcommand{\svaNImpl}{\mathbin{|{\mkern-3mu}\rightarrow}}  
\newcommand{\svaOImpl}{\mathbin{|{\mkern-3mu}\Rightarrow}}  
\newcommand{\svaConcat}[1]{\mathbin{\#\mkern-2mu\#}#1}      

\algrenewcommand\algorithmicrequire{\textbf{Input:}}
\algrenewcommand\algorithmicensure{\textbf{Output:}}

\usepackage{textcomp}

\DeclareCaptionFormat{listing}{\rule{\columnwidth}{0.1pt}\vskip1pt#1#2#3}
\newcommand{\tb}{\mathbf{B}}

\newcommand{\spec}{\mathbf{S}} 
\newcommand{\design}{\mathbf{D}} 
\newcommand{\pdesign}{\mathbf{D}_p}
\newcommand{\signalset}{\mathbb A} 
\newcommand{\speccnf}{\phi_{\spec}} 
\newcommand{\heuset}{\prec_{\mathbb{C}}} 
\newcommand{\heulitset}{\prec_{\mathbb{C}}'} 
\newcommand{\unrolleddesignsmt}{\Phi^{\tb}} 
\newcommand{\designcnf}{\phi_{\design}} 
\newcommand{\var}{\mathcal{V}} 
\newcommand{\set}[1]{\mathcal{#1}} 

\newcommand{\depth}{\mathsf{Dep}} 
\newcommand{\mapd}{M_{\signalset \to \designcnf}}
\newcolumntype{C}[1]{>{\centering\let\newline\\\arraybackslash\hspace{0pt}}m{#1}}
\newcommand{\secret}[1]{\llbracket #1 \rrbracket}
\usepackage{caption}
\usepackage{titlesec}
\titlespacing*{\section}
{0pt}{0.2\baselineskip}{0.2\baselineskip}
\titlespacing*{\subsection}
{0pt}{0.2\baselineskip}{0.2\baselineskip}
\usepackage{subcaption}
\usepackage{etoolbox}

\usepackage{pifont}

\usetikzlibrary{arrows,automata,circuits.logic.US,calc,3d,backgrounds}

\usepackage{tikz-3dplot}
\usepackage{pgfplots}
\usetikzlibrary{shapes,arrows,patterns}
\usetikzlibrary{pgfplots.statistics}
\usepackage{siunitx}

\definecolor{CommentGreen}{rgb}{0,0.6,0}
\definecolor{numbering}{gray}{0.5}
\definecolor{keywordcolor}{rgb}{0.63,0,0.42}
\definecolor{string}{rgb}{1,0,0}
\definecolor{babypink}{rgb}{0.96, 0.76, 0.76}
\definecolor{azure(web)(azuremist)}{rgb}{0.94, 1.0, 1.0}

\lstdefinestyle{flang} {
    basicstyle=\ttfamily\footnotesize,
    numbers=none,
    comment=[l]{//},
    fontadjust=true,
    basewidth=0.5em,
    keywordstyle={\bfseries\color{magenta}},
    commentstyle=\color{CommentGreen},
    stringstyle=\color{blue},
    tabsize=4,
    morekeywords={[1] always, posedge, begin, if, end
    }
}

\definecolor{codegreen}{rgb}{0,0.6,0}
\definecolor{codegray}{rgb}{0.5,0.5,0.5}
\definecolor{codepurple}{rgb}{0.58,0,0.82}
\definecolor{backcolour}{rgb}{0.95,0.95,0.92}

\lstdefinestyle{motivating}{
    language=Verilog,
    backgroundcolor=\color{backcolour},   
    commentstyle=\color{codegreen},
    keywordstyle=\color{magenta},
    numberstyle=\tiny\color{codegray},
    stringstyle=\color{codepurple},
    basicstyle=\ttfamily\footnotesize,
    breakatwhitespace=false,         
    breaklines=true,                 
    captionpos=b,                    
    keepspaces=true,                 
    numbers=left,                    
    numbersep=5pt,                  
    showspaces=false,                
    showstringspaces=false,
    showtabs=true,                  
    tabsize=2,
    morekeywords={assert,property,cover, set, info, declare, fun, check, sat }
}

\lstdefinestyle{vcode}{
	language=Verilog,
    commentstyle=\color{CommentGreen},
	stringstyle=\color{string},
	keywordstyle=\bfseries\color{magenta},
	basicstyle=\scriptsize\ttfamily,
	numbersep=5pt,
	frame=lines,
	breaklines=true,
	showstringspaces=false,
	tabsize=2,
    morekeywords={assert,property,cover, set, info, declare, fun, check, sat}
}

\lstdefinestyle{dimcas}{
	language=Verilog,
    commentstyle=\color{CommentGreen},
        morecomment=[f]{//}, 
	stringstyle=\color{string},
	keywordstyle=\bfseries\color{magenta},
	basicstyle=\footnotesize\ttfamily,
        escapeinside={<@}{@>},
	frame=single,
	breaklines=true,
	showstringspaces=false,
	tabsize=2,
    xleftmargin=2em,
    morekeywords={assert,property,cover, set, info, declare, fun, check, sat }
}

\lstdefinestyle{shcode}{
    language=sh,                        
    commentstyle=\color{comment},       
    stringstyle=\color{string},        
    keywordstyle=\bfseries\color{keyword}, 
    basicstyle=\footnotesize\ttfamily, 
    captionpos=b,                     
    numbers=left,             
    numberstyle=\tiny,                
    frame=single,                    
    showstringspaces=false,  
    tabsize=2,                          
    xleftmargin=2em,                
    showstringspaces=false             
}

\lstdefinestyle{smtlib2}{
        comment=[l]{//},
        commentstyle=\color{CommentGreen},
	stringstyle=\color{string},
	keywordstyle=\bfseries\color{magenta},
	basicstyle=\footnotesize\ttfamily,
	numbersep=5pt,
	frame=lines,
	breaklines=true,
	showstringspaces=false,
	tabsize=2,
    xleftmargin=2em,
    morekeywords={assert,property,cover, set, info, declare, fun, check, sat }
}
 
\newcommand{\signalname}[1]{\texttt{#1}}
\newcommand{\randomize}[1]{ \widetilde{#1}}

\newcommand{\liu}[1]{\textcolor{brown}{[Zhaoxiang: #1]}}
\newcommand{\tool}{BlindMarket\xspace}

\newcounter{myexample}

\newenvironment{codeexample}[1]
{%
\begin{figure}[!h]%
\newcommand\ColumnFigureCaption{{#1}}%
\refstepcounter{example}
}
{%

\caption*{{EXAMPLE~\theexample: }\ColumnFigureCaption}%
\end{figure}%
}

\newcounter{example}

\author{
\IEEEauthorblockN{
Zhaoxiang Liu\IEEEauthorrefmark{1},
Samuel Judson\IEEEauthorrefmark{2},
Raj Dutta\IEEEauthorrefmark{3},
Mark Santolucito\IEEEauthorrefmark{4},
Xiaolong Guo\IEEEauthorrefmark{1},
Ning Luo\IEEEauthorrefmark{5}
}

\IEEEauthorblockA{\IEEEauthorrefmark{1}Kansas State University
\{zxliu, guoxiaolong\}@ksu.edu}

\IEEEauthorblockA{\IEEEauthorrefmark{2}Trail of Bits
samuel.judson@trailofbits.com}

\IEEEauthorblockA{\IEEEauthorrefmark{3}Silicon Assurance
rajgautamdutta@siliconassurance.com}

\IEEEauthorblockA{\IEEEauthorrefmark{4}Barnard College, Columbia University
msantolu@barnard.edu}

\IEEEauthorblockA{\IEEEauthorrefmark{5}University of Illinois Urbana-Champaign
nl27@illinois.edu}
}

\begin{document}

\title{\tool: Enabling Verifiable, Confidential, and Traceable IP Core Distribution in Zero-Trust Settings}


\maketitle
\begin{abstract}

We present \tool{}, an end-to-end zero-trust distribution framework for hardware IP cores that operates without relying on any trusted third party. \tool{} allows two parties, the IP user and the IP vendor, to complete an IP trading process with strong guarantees of verifiability and confidentiality before the transaction, and then traceability after. 
We propose verification heuristics and adapt the cone of influence-based design pruning to overcome the limited scalability common to cryptographic protocols and the hardness of the underlying hardware verification.
We systematically evaluate our framework on a diverse set of real-world hardware benchmarks. The results demonstrate that BlindMarket successfully verifies 12 out of 13 IP designs and achieves substantial performance improvements enabled by design pruning and control-flow guided heuristics.

\end{abstract}


\input{sections/introduction}

\input{sections/prelims}
\section{Related Work}\label{sec:relat}
\input{sections/related}

\input{sections/threatmodel}
\input{sections/methodology} 
\input{sections/optimization}

\input{sections/experiments} 



\input{sections/conclusion} 
\input{sections/ack}


\bibliographystyle{IEEEtran}
\bibliography{ref}

\end{document}

%% file: sections/introduction.tex
\let\thefootnote\relax\footnotetext{\IEEEauthorrefmark{2} Work completed while at Yale University.}

\section{Introduction}\label{sec:intro}
The integration of third-party intellectual property (IP) cores into system-on-chip (SoC) designs is now a standard practice in semiconductor development. 
By 2021, about 1.15 trillion devices had been shipped with chips containing various IP cores, highlighting the market’s scale~\cite{semiconductor2022global}.
In this market, \emph{IP vendors} such as Arm, Intel, Cadence, and Synopsys provide the IP core design together with supplementary documentation and test benches to the \emph{IP user}. Their revenue primarily comes from licensing these high-performance hardware designs and charging a fee for each chip manufactured that incorporates the IP.



Formal verification enables IP users to exhaustively verify design properties before purchase. SAT-based hardware verification stands out due to recent breakthroughs in SAT solving~\cite{fern2017detecting,ceesay2024mucfi,guo2017pch,zhang2018end,meng2021rtl,zhang2022gate}.
However, the nature of formal verification requires the exposure of the IP core to the IP user so that the latter can undertake the verification, which raises significant risks of IP theft~\cite{reuters_asml_iptheft_2024,cnn_tesla_xiaopeng_2019, halbert2016intellectual,mishra2017hardware}. 
The IP vendor is at risk of IP theft and commercial loss without strict protections when they provide temporary access to their products and disclose detailed product information prior to a transaction closing. 
Privacy leakage is, however, bidirectional. An untrusted IP vendor can exploit the specifications to infer the IP user’s design needs and purchasing intentions. Such knowledge not only exposes system details of value to competitors~\cite{ray2023samsung}, but also enables the vendor to manipulate pricing, posing additional commercial risks.


Existing IP core protection methods focus on preventing illegal hardware copying through techniques such as obfuscation and watermarking~\cite{kahng2001constraint, alkabani2007active, kamali2022advances, chakraborty2009harpoon}. This line of work is orthogonal to ours, which aims to preserve confidentiality throughout the IP verification phase. While legitimate efforts to shield hardware IP, such as the IEEE 1735 standard~\cite{7274481}, can help, they have been proven insufficient to prevent theft~\cite{speith2022not}, leading to financial losses across the industry. Recent demands for \emph{zero-trust} approaches highlight the need for a secure framework that enables IP exchange without reliance on trusted third parties~\cite{7274481}.

There remains a strong need for IP core traceability, which allows users to identify the origin of an IP and avoid redundant verification. The risks of untraceable hardware gained renewed attention in 2024 during the Israel–Hezbollah conflict, where explosive-laden pagers were falsely labeled as Gold Apollo products~\cite{reuters}, underscoring the security implications of missing traceability in critical systems.

In sum, an ideal IP core distribution platform should achieve the following three goals, all without a trusted party:
 
\noindent
{\bf Verifiability.} The platform should guarantee that an acquired IP core behaves as advertised. To this end, the IP user must be able to perform IP verification before entering an agreement with the vendor.

\noindent
{\bf Confidentiality.} The IP vendor's proprietary designs and the IP user's sensitive data should both remain private until a legal framework is established, enabling integration. 

\noindent
{\bf Traceability.} The platform should offer verifiable provenance and integrity checks for IPs, ensuring that distributed IPs are authentic while remaining agnostic to post-acquisition use.

Achieving verifiability, confidentiality, and traceability simultaneously in hardware IP distribution is challenging due to inherent tensions among these goals. Verifiability often requires exposing design details to the user, compromising confidentiality. Likewise, embedding traceability markers can alter control or data paths, potentially breaking logic equivalence checking.
To mitigate such tensions, a natural direction is to leverage cryptographic techniques for verification. \texttt{ppSAT}~\cite{luo2022ppsat} was originally designed to protect the privacy of Boolean satisfiability queries, and its techniques, though applicable in principle, have not been applied to hardware and do not provide an end-to-end zero-trust workflow for real-world IP trading. Its limited scalability further restricts its practicality for hardware IP verification.

To address the above limitations, we propose \tool. Our contributions are summarized as follows.
\begin{itemize}
\item We propose an end-to-end ``zero-trust" framework, \tool, that allows the IP user and the vendor to establish a trusted IP trading process to privately select and verify IP cores of interest before the transaction takes place without relying on any trusted third-party.

\item We turn the idea of privacy-preserving verification into a practical end-to-end hardware verification framework with tailored formula generation and hardware-focused optimizations.

\item We design a cryptographic ledger–based IP ownership auditing mechanism that complements \tool{} by enabling post-trade validation of provenance and ownership. This mechanism provides verifiable traceability of IP cores across transactions.

\item We implement our framework and evaluate it on real-world IP cores. The results demonstrate that our approach efficiently verifies functional properties, achieving performance improvements in 12 out of 13 benchmarks. Its effectiveness is further illustrated in Figure~\ref{fig:overall}.

\end{itemize}

%% file: sections/prelims.tex
\section{Preliminaries}\label{sec:pre}

\subsection{SAT-based Hardware Verification}
Many hardware verification tasks can be reduced to SAT problems. 
The hardware design can be modeled as a transition system, and its behavior over a bounded number of time steps can be encoded as a Boolean formula \( \phi_D \) by unrolling the transition relation across multiple cycles.


The property to be verified in a hardware design is expressed as a Boolean formula \( \varphi \). In the open-source hardware community, temporal properties are of much interest. In this paper, we focus on the property using the operator of 
Overlap Implication (OI), Non-Overlap Implication (NOI), and Concatenation (Concat)~\cite{dobis2024formal}. 
Together, these operators provide a composable basis for expressing temporal relationships over bounded execution traces.
Specifically, OI (\(\mathrm{A} \svaNImpl \mathrm{B}\)) asserts that whenever A holds, B must hold at the same clock cycle, whereas NOI (\(\mathrm{A} \svaOImpl \mathrm{B}\)) requires B to hold in the next clock tick if A holds in the current clock tick. The Concat operator (\(\mathrm{A} \svaConcat{\mathrm{N}} \mathrm{B}\)) checks that A must be true on the current clock tick and B on the N$^{\text{th}}$ clock tick. Here, A and B are boolean expressions defined as a sequence~\cite{vijayaraghavan2005practical}. One typically constructs the conjunction \( \phi_D \land \neg \varphi \) in conjunctive normal form (CNF). If this formula is satisfiable, it reveals a counterexample that violates the intended property; otherwise, the property holds within the given bound.



\subsection{Cryptographic Tools}\label{subsec:overviewppsat}


\paragraph{Privacy-Preserving SAT Solving}

Secure Two-Party Computation (2PC) is a cryptographic protocol that enables two parties with private inputs to jointly compute a function $f(a,b)$ without revealing anything beyond the final output. 

Privacy-preserving SAT solving (\texttt{ppSAT})~\cite{luo2022ppsat} leverages 2PC to allow two parties to check the satisfiability of the conjunction of their CNF formulas without revealing the formulas themselves.
In \texttt{ppSAT}, each party encodes its CNF formula as a pair of matrices. Let $\phi = \bigwedge_{j=1}^m C_j$ denote a CNF formula with $m$ clauses over $n$ variables. Each clause $C_j$ is represented by a pair of binary vectors $(P[j], N[j])$ of length $n$, where the encoding is defined as follows:
\begin{equation*} (P[j][i], N[j][i]) = \begin{cases} (1, 0) & \text{if $x_i$ appears in $C_j$,}
\\ (0, 1) & \text{if $\neg x_i$ appears in $C_j$},\\  (0, 0) & \text{otherwise}. \end{cases} \end{equation*}

\NiceMatrixOptions{code-for-first-row = \color{red},
 code-for-first-col = \color{red},
 code-for-last-row = \color{red},
 code-for-last-col = \color{red}}


The complete formula $\phi$ is encoded as a pair of binary matrices $(P, N) \in {0,1}^{n \times m}$, where each column $j$ corresponds to the clause representation $(P[j], N[j])$. 

We consider \texttt{ppSAT} as consisting of two phases: the {\bf sharing} (\texttt{ppSAT.share}) phase and the {\bf solving} (\texttt{ppSAT.solving}) phase.  Each party first encodes its private Boolean formula as a pair of matrices, denoted by \( (P_a, N_a) \) and \( (P_b, N_b) \), respectively. 
In the sharing phase, the two parties secret-share the concatenated matrices using Yao’s sharing~\cite{yao1986generate}. A value \( v \) is said to be \emph{Yao-shared}, denoted by \( \secret{v} \), if it is the output of a GC-based 2PC computation and has not been revealed to either party. 
In the subsequent solving phase, \texttt{ppSAT} performs SAT solving by updating the shared matrix representation within the 2PC framework.  Each update is referred to as a \emph{giant step}, which consists of three procedures: \emph{unit propagation}, \emph{pure literal elimination}, and \emph{variable decision branching}. In particular, \emph{variable decision branching} relies on an external heuristic to determine which variable assignment to try when no variable can be directly decided, and the solving process must proceed by branching.

\label{sec:crypto}
\noindent
\paragraph{Oblivious transfer.}
Oblivious transfer (OT) is a fundamental cryptographic primitive that enables a sender to transfer one of many pieces of information to a receiver, while preserving the privacy of both parties. In a \emph{1-out-of-$N$ OT} protocol, the sender holds a vector of messages $(m_1, m_2, \ldots, m_N)$, and the receiver holds an index $\alpha \in \{1, \ldots, N\}$. At the end of the protocol, the receiver learns $m_\alpha$ and nothing about $\{m_j\}_{j \ne \alpha}$ while the sender learns nothing about the receiver’s choice $\alpha$. We denote this invocation as
\[(m_i, \bot )\leftarrow {\sf OT}(i, \{m_1, m_2, \cdots, m_N \}) \]

\noindent
\paragraph{Ledger System.}
A ledger is a record-keeping system that tracks the state and ownership of assets. A cryptographic ledger extends this concept with cryptographic primitives to provide verifiable storage and consensus. \tool requires such a ledger for transparency but is agnostic to the underlying consensus mechanism (centralised, decentralised, permissionless, \emph{etc.}), as long as it supports reliable storage of cryptographically validated records and the simple queries defined below.

%% file: sections/related.tex




We compare our work against prior efforts from three perspectives: verifiability, confidentiality, and traceability.

\begin{table}[h]
    \centering
        \caption{Existing methods cannot provide privacy protection for formal-based IP verification without a third party.}
    \begin{tabularx}{\columnwidth}{Xccc}
    \toprule
    \textbf{Approach} & \textbf{Parties} & \textbf{Verification} & \textbf{Stage}\\
    \midrule
    TIPP~\cite{kahng2001constraint,alkabani2007active,kamali2022advances} &- & - &Post-fabrication \\
    IEEE 1735~\cite{7274481} & 3& Simulation/Synthesis &Pre-silicon \\
    Garbled EDA~\cite{hashemi2022garbled} & 3 & Simulation& Pre-silicon (Netlist)\\
    MP$\ell$$\circ$C~\cite{mouris2023mploc} &2 & Simulation & Pre-silicon (Netlist)\\
        Pythia~\cite{mouris2020pythia} &2 & Simulation & Pre-silicon (Netlist)\\
    \textbf{This Work} & \textbf{2} &\textbf{Formal Logic}  & \textbf{Pre-silicon(RTL)}\\
    \bottomrule
    \end{tabularx}
    \label{tab:quantitativecomp}
\end{table}
\input{revision/related}

\noindent {\bf Traceability.}
Blockchain can enhance hardware security by tracking and authenticating ICs across supply chains. Islam et al.~\cite{islam2019enabling} used smart contracts to log IC identity, ownership, and PUF-derived data, while Chaudhary et al.~\cite{chaudhary2021auto} reduced on-chain storage by moving CRPs off-chain with on-chain references. ICToken~\cite{balla2024ictoken} extends tracking from fabrication to end users via NFT tokenization~\cite{wang2021non} and logic locking. However, existing approaches primarily focus on post-fabrication traceability and provide limited coverage in earlier design stages. Our work fills this gap by integrating soft IP traceability into a blockchain to enable secure and transparent lifecycle management.

%% file: revision/related.tex
{\bf Verifiability \& Confidentiality.}
As summarized in Table~\ref{tab:quantitativecomp}, existing approaches either lack verification support~\cite{kahng2001constraint, alkabani2007active, kamali2022advances, chakraborty2009harpoon}, rely on simulation-based verification~\cite{mouris2023mploc,mouris2020pythia}, or require a trusted third party~\cite{7274481,hashemi2022garbled}. Whereas our work enables privacy-preserving formal verification for a wide range of IP designs in a two-party setting.

Various IP protection techniques, including watermarking, active metering, and logic locking~\cite{kahng2001constraint, alkabani2007active, kamali2022advances, chakraborty2009harpoon}, as summarized in Table~\ref{tab:quantitativecomp}, aim to deter IP misuse. However, these approaches primarily protect IP after trading or fabrication and provide no guarantees of verifiability. The IEEE 1735 standard~\cite{7274481} provides an encryption and management framework involving a trusted EDA tool vendor, yet attacks~\cite{speith2022not, chhotaray2017standardizing} have exposed inherent weaknesses in its deployment within current commercial EDA tools. Garbled EDA~\cite{hashemi2022garbled} enables an IP user to obtain simulation results without revealing sensitive information such as the process design key, IP core, or EDA binaries, but it relies on a trusted EDA tool provider. MP$\ell$$\circ$C~\cite{mouris2023mploc} combines secure multiparty computation (MPC) and logic locking to support privacy-preserving IP simulation, while Pythia~\cite{mouris2020pythia} allows the IP vendor to generate a zero-knowledge proof demonstrating the correctness of the computation to the IP user. As reflected in Table~\ref{tab:quantitativecomp}, these methods are limited to simulation-based verification at the netlist level and are primarily evaluated on the ISCAS'85 benchmarks, leaving open questions regarding scalability and applicability to formal logic verification.

%% file: sections/threatmodel.tex
\section{Threat Model}
\label{sec:threat}
\input{revision/threatmodel}In the \textit{IP distribution phase}, a dishonest IP user may actively deviate from the protocol to obtain license authorization without approval from the IP vendor fraudulently. Such adversaries can collude with compromised authorized users to forge a license for protected IP content.

%% file: revision/threatmodel.tex
BlindMarket operates in a zero-trust setting without relying on any trusted broker or verification provider. The system consists of two phases: (1) secure verification and (2) IP distribution, and its security relies solely on cryptographic protocols and on-chain records. During the secure verification phase, we adopt the semi-honest adversary model, where both parties follow the protocol but may attempt to infer additional information from the transcript. Under this model, the IP user learns only the final verification result and public metadata, while the IP vendor learns neither the user’s selected IP index nor the property specification.
A malicious party may abort the protocol or provide malformed inputs, which may affect availability or correctness. Preventing such behaviors would require malicious-secure computation with input-consistency checks, which is beyond the scope of this work.

%% file: sections/methodology.tex
\section{\tool System Design }\label{sec:med}

\subsection{\tool Workflow}\label{subsec:workflow}
\tool involves two parties: the IP user and the IP vendor. In practice, an IP vendor typically maintains a portfolio of IP designs available for licensing. We model this portfolio as a set of abstract specifications, each describing the functionality of a design without disclosing implementation details or internal structures. Based on these high-level descriptions, an IP user selects a target design to verify while keeping their selection and property of interest hidden from the vendor. One of the core challenges is to allow the user to verify the correctness of a selected design without revealing the design itself or the user’s verification intent. 

In addition to confidentiality and verifiability, \tool enforces traceability after the IP core is delivered. Specifically, the vendor issues a usage certificate bound to the authorized user. This certificate enables the user to incorporate the IP into their design. \tool ensures that even in the presence of collusion, an authorized user cannot transfer or replicate this certificate for the unauthorized user. 

\begin{figure}[t]
    \centering
    \includegraphics[width=\columnwidth]
    {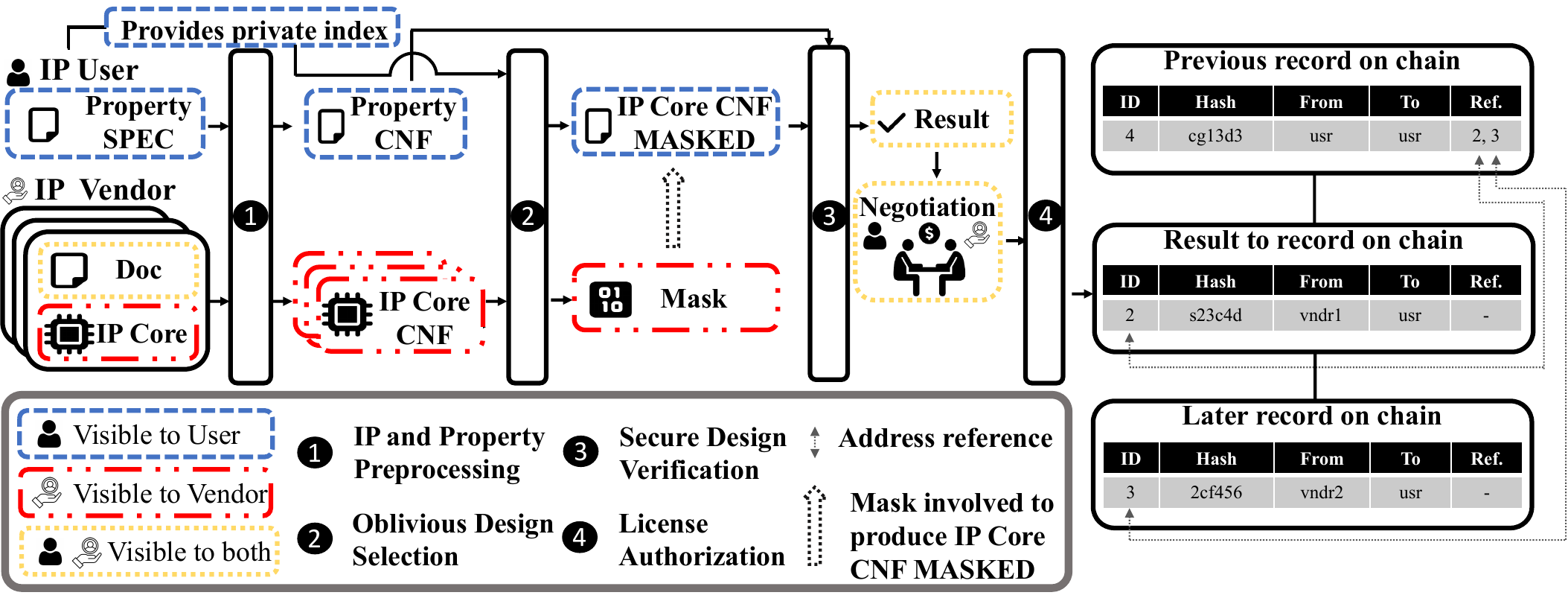}
    \caption{ End-to-end workflow of secure IP verification and licensing in \tool, organized into four phases. (1) In the preprocessing phase, the vendor encodes each IP core into CNF and masks it using randomness, while the user encodes the verification property. (2) During oblivious design selection, the user retrieves the masked CNF of the selected design via 1-out-of-$N$ OT, without revealing the selection to the vendor. (3) In the secure design verification phase, the vendor and user engage in a 2PC protocol to reconstruct the selected design and jointly verify it against the property using \texttt{hw-ppSAT}. (4) Finally, if verification succeeds, the vendor issues a certificate bound to the authorized user and selected IP, publishing it to the ledger for traceability.}
    \label{fig:overview}
\end{figure}

To achieve these goals, \tool consists of four core phases: (1) IP and property preprocessing, where both the user and the vendor convert IP designs and property to appropriate formats; (2) oblivious design selection, where the user selects a design without learning its content or revealing their selection; (3) secure design verification, allowing the user to check correctness without accessing the full design; and (4) license authorization, where the vendor issues a certificate that binds the authorized user to the selected IP.
Figure~\ref{fig:overview} summarizes the process. 



\noindent{\bf Customized IP and property translations.}
For each commercial IP product, the vendor discloses a high-level functional specification and a set of \textit{observable signals} to the IP user. The set of observable signals typically includes the primary I/O and a subset of internal signals. To support secure verification, the vendor locally converts the design into a Boolean formula and provides \textit{semantic map} of the observable signals, which are detailed in Section~\ref{subsec:preprocess}.

The IP vendor then reveals the \textit{semantic map} to the IP user so that they can model the property as a Boolean formula using variables with the same semantics as the vendor does. The value of IP usually lies in the implementation of logical operations and interconnection, which is not shared in \tool. To balance efficiency and confidentiality, \tool provides the optimization approach for design pruning when the user is willing to reveal assertion signals. We detail the optimization in Section~\ref{subsec:prune}, and our evaluation results (see Section~\ref{exp:hwppsat}).

\noindent{\bf Oblivious design selection.}
The IP user selects a design of interest from the vendor’s portfolio, while keeping the selection hidden to protect purchase intentions. This phase enables private design selection for subsequent verification without revealing the chosen design to either party.

After preprocessing, the Boolean formula of all designs in the vendor portfolio are obfuscated with the same randomness. The obfuscated formula is then transmitted via an oblivious transfer (OT) protocol. The way we obfuscate the formula is information-theoretically secure, and therefore ensures confidentiality of the design. Meanwhile, the security of OT guarantees the privacy of the IP users' choice and thus interests. The technical details of the protocol are presented in Section~\ref{subsec:ot_selection}.

\noindent{\bf Secure design verification.}
The secure design verification protocol is initialized after the user obtains the obfuscated design, as detailed in Section~\ref{subsec:2pc}. 
In this phase, the user provides the obfuscated Boolean formula of the design and the Boolean formula of their property. At the same time, the vendor inputs the randomness he uses to mask all formulas. The protocol first demasks the design formula using input from two parties in 2PC. The demasked design and the property are then checked for satisfiability using a subsequent privacy-preserving SAT-solving protocol. The output of the solver is revealed to users, indicating that the design satisfies the expected property.

\noindent{\bf License authorization.}
The vendor registers each design in a ledger system using a unique identifier. To support a zero-trust framework, we implement the ledger using a blockchain-based infrastructure. Upon the completion of a sale, the vendor authenticates the transaction by appending a new entry to the ledger. During integration, the IP vendor uploads an additional record that references the previously recorded purchase. The structure of the record and operations are described in detail in Section~\ref{subsec:ip-traceability}.

\subsection{Customized IP and Property Translations}\label{subsec:preprocess}

Existing translation tools built on hardware synthesis frameworks typically bind the design to a specific cell library and flatten it into gate-level netlists. 
Such flattening often removes high-level semantic structures, making it difficult to express verification properties that are naturally defined at the behavioral level. 
Moreover, complex operators (e.g., multiplexers) are decomposed into primitive logic gates, which introduces a large number of auxiliary variables and clauses during CNF generation. 
Since \tool encodes the CNF as a pair of binary matrices $(P, N) \in \{0,1\}^{n \times m}$, larger values of $n$ and $m$ directly increase the computational and communication overhead of the subsequent secure protocols.
Therefore, we customize a word-level translation for IP designs and properties that maintains a small number of variables, and both users and vendors independently encode the design and properties into Boolean formulas based on a publicly agreed-upon bound.

\subsubsection{Formula Generation of Design}\label{subsec:ipdesign}
On the IP vendor side, the design is encoded as a Boolean formula through the following steps.  
Our static analysis frontend (based on~\cite{takamaeda2015pyverilog,wolf2016yosys}) takes the synthesizable RTL as input and emits a word-level intermediate representation, which is then translated into a quantifier-free bit-vector SMT formula (QF-BV)~\cite{barrett2010smt}, where each variable is modelled as a \emph{fixed-size bit vector}.  
For instance, in Example~\ref{fig:rtlcounter}, the signal \signalname{counter}, defined starting at line~3, is translated into its corresponding SMT bit-vector representation beginning at line~10. 
The resulting SMT formula is then converted into CNF via \emph{bit-blasting}~\cite{barrett1998decision} and \emph{Tseytin transformation}~\cite{tseitin1983complexity}.  
Bit-blasting translates bit-vector operations into Boolean logic, while the Tseytin transformation encodes Boolean formulas in CNF using auxiliary variables without exponential growth. 


Meanwhile, \tool frontend allows the vendor to automatically construct a \textit{semantic map} for the \textit{observable signals}, which correspond to a small subset of design variables. The semantic map assigns each observable signal a unique literal, as shown in lines~15–16, where each bit of the variables \signalname{counter} and \signalname{reset} is assigned a unique literal. This mapping enables the IP user to formulate properties in Section~\ref{subsec:ipspec}.
The signals used in the control expressions defined as control signals are also tracked in the semantic map (such as \texttt{reset} in Example~\ref{fig:rtlcounter} line~2), which remains private and is weighted by its depth in its statement's abstract syntax tree (AST)~\cite{alfred2007compilers}, We leverage this metric to accelerate the SAT solver, as detailed in Section~\ref{sec:heuristics}.
{Example}~\ref{fig:rtlcounter} demonstrates how a simple counter updates \signalname{counter} based on previous value of control signal \signalname{reset} and \signalname{enable} symbolically.
The SMT constraints in lines 10–13 specify an equivalent logical formula that encodes its behaviour, where we append $\#i$ to each signal to denote the timestamp. Specifically, \signalname{counter\#2} denotes the symbolic value of \signalname{counter} at the second clock cycle and is updated by a nested ternary expression referencing \signalname{counter\#1} from the first clock cycle.
Furthermore, the vendor provides a semantic map that associates the 3-bit signal \signalname{counter\#2} with the literal $\{\ell_{4},\ell_{3},\ell_{2}\}$ and the 1-bit signal \signalname{reset\#1} with $\ell_1$, while also recording the depths of \signalname{reset\#1} and \signalname{enable\#1} as 1 and 2, respectively.

\begin{codeexample}{Incrementing counter with synchronous reset}
\begin{lstlisting}[style=motivating,basicstyle=\ttfamily\scriptsize]
  always @(posedge clk) begin
    if (reset)
      counter <= 3'b0; // Reset the counter to 0
    else if (enable)
      counter <= counter + 3'b10; // Increment the counter by 2
    else
      counter <= counter + 3'b1; // Increment the counter by 1
  end
/* SMT Formula */
counter#2 <--> (ite (= |reset#1| #b1)
#x00(ite (= |enable#1| #b1)
    (bvadd |counter#1| #x02)
    (bvadd |counter#1| #x01)))
/* Semantic map */
{reset#1 : 1, counter#2[0] : 2, 
counter#2[1] : 3, counter#2[2] : 4}
/*Ctrl Depth*/
{reset#1 : 1, enable#1 : 2}

\end{lstlisting}
\label{fig:rtlcounter}
\end{codeexample}






\subsubsection{Boolean Encodings of Properties }\label{subsec:ipspec}
The IP user then applies the \textit{semantic map} to encode its property as a Boolean formula. 
\tool focuses on properties composed of three common primitives, namely NOI, OI, and Concat, each of which can be translated into basic Boolean logic.
NOI is represented by $\neg \signalname{a}_{i} \vee \signalname{b}_{i+1}$, indicating that if signal $\signalname{a}$ holds at time step $i$, then $\signalname{b}$ must hold at the next time step $i+1$. In contrast, OI is encoded as $\neg \signalname{a}_{i} \vee \signalname{b}_{i}$, requiring that $\signalname{b}$ hold in the same cycle as $\signalname{a}$. Concat expresses a sequential pattern and is encoded as $\signalname{a}_{i} \wedge \signalname{b}_{i+n}$, meaning that signal $\signalname{a}$ must hold at time $i$, and signal $\signalname{b}$ must hold exactly $n$ cycles later.

In Example~\ref{fig:rtlcounter}, the user obtains the semantic map from the IP vendor, and a property such as
$
\signalname{reset\#1} \rightarrow \bigl(\signalname{counter\#2} = 0\bigr)
$,
can be used to check whether \signalname{counter} is reset to zero whenever \signalname{reset} is asserted. Under the semantic map, this requirement is translated into the Boolean formula 
$\lnot \ell_1 \lor (\neg \ell_2 \wedge \neg \ell_3 \wedge \neg \ell_4)$.

\subsection{Oblivious Design Selection}\label{subsec:ot_selection}
Oblivious selection is performed prior to secure verification to prevent the IP vendor from knowing the IP design of the user's interests. 
In standard verification workflows, the user must disclose the target they wish to verify to initiate the process. Such disclosure can be problematic, particularly when the vendor may exploit knowledge of user preferences to manipulate pricing or restrict offerings.

\tool leverages oblivious transfer (OT) to address this challenge. OT enables an IP user (the receiver) to obtain one design from a set offered by the vendor (the sender) without revealing which design is selected. However, directly applying OT will compromise confidentiality by exposing the full IP design to the user. To preserve design privacy while enabling verification, the OT procedure must be integrated with the privacy-preserving framework. This integration requires the SAT solver to operate in a white-box manner, as detailed in Section~\ref{subsec:2pc}.

Initially, the CNF of each design $\phi_k$ in the design portfolio $\Phi$ is encoded as a pair of binary matrices $({P}_k, {N}_k)$ on the vendor side. Then each pair $({P}_k, {N}_k)$ is masked using a shared pair of random binary matrices $({R}_P, {R}_N)$. We require $({R}_P, {R}_N)$ to be refreshed for every verification session. Otherwise, colluding users could XOR masked instances obtained from different runs to cancel the masks and infer structural relationships between protected designs. Specifically, the masked representation is computed as $({\randomize{{P}_k}}, \randomize{{N}_k}) = ({P}_k \oplus {R}_P, {N}_k \oplus {R}_N)$. The collection of masked matrices ${({\randomize{{P}_k}}, \randomize{{N}_k})}$ is used as the vendor’s input to the oblivious transfer (OT) protocol. 

On the IP user side, the user selects an index $i$ corresponding to the desired design based on the public metadata provided in the vendor's design portfolio. This index serves as the receiver's input to a 1-out-of-$N$ OT protocol. During the OT execution, the two parties jointly run the protocol such that, at the conclusion, the user obtains the pair of masked matrices $(\randomize{P}_i, \randomize{N}_i)$ corresponding to the selected design. The IP vendor retains the random masking matrices $(R_P, R_N)$.





\begin{figure}[t]
    \centering
    \includegraphics[width=\columnwidth]{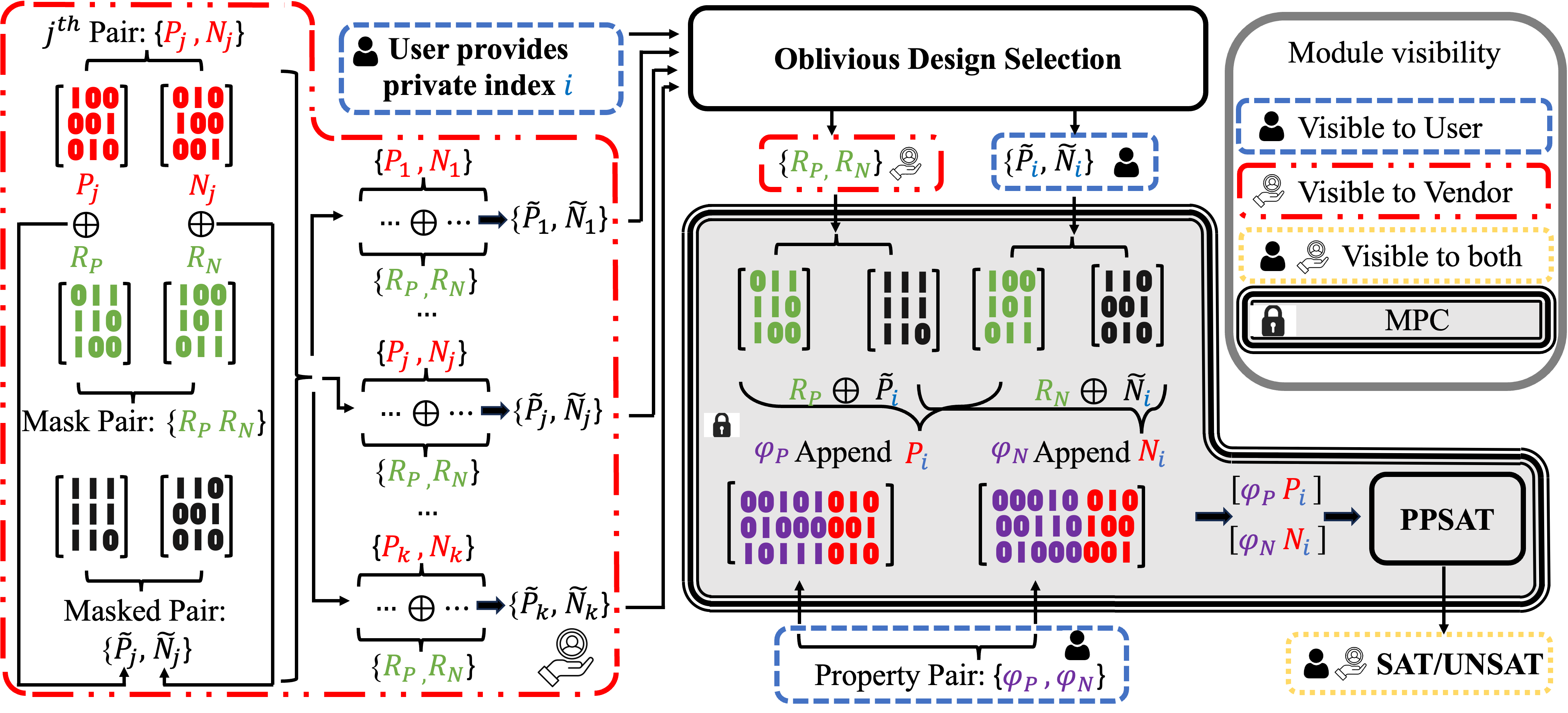}
    \caption{Secure hardware design verification workflow in \tool. The vendor masks each CNF-encoded design using random matrices, and the user selects a target design via oblivious transfer (OT), receiving only the masked version. The vendor and user then run a 2PC protocol to reconstruct the selected design and combine it with the user’s property to form $\psi = \phi_i \land \neg\varphi$. The satisfiability of $\psi$ is evaluated using \texttt{hw-ppSAT} under 2PC, revealing only the final result to both parties while preserving mutual confidentiality.}
    \label{fig:secure-verification}
\end{figure}

\subsection{Secure Design Verification}
\label{subsec:2pc}
Secure design verification aims to determine whether a proprietary design satisfies a user-specified property, without revealing the design to the user or disclosing the property to the vendor. 
However, applying \texttt{ppSAT} directly is insufficient in our setting as it requires both parties to agree on a common verification target in the setup. This synchronization step reveals the IP user's selection to the vendor.

To resolve this issue, after the oblivious design selection protocol, the user obtains a pair of obfuscated matrices $(\randomize{P}_i, \randomize{N}_i)$ corresponding to the selection index $i$. Following this, \tool initiates a secure 2PC protocol in which the user inputs the masked design matrices and the property $\varphi$, while the vendor inputs the retained random masking shares $(R_P, R_N)$. The design matrices are then reconstructed by demasking in 2PC:
\begin{equation}
\label{eq:deobfuscate}
(P_i, N_i) = (\randomize{P}_i \oplus R_P,\, \randomize{N}_i \oplus R_N).
\end{equation}

This demasked pair of matrices encodes the selected design $\phi_i$. The verification task is then formulated as the satisfiability of the Boolean formula $\psi = \phi_i \land \neg\varphi$, which is secret-shared between the two parties within the 2PC. The \texttt{ppSAT.Solve} protocol is then invoked to evaluate the satisfiability of $\psi$ and the satisfiability result is revealed to both parties at the end.
Throughout this process, the IP design remains hidden from the user, and the verification property remains unknown to the vendor, ensuring mutual confidentiality.


\subsection{\texttt{hw-ppSAT}}\label{sec:heuristics}



Numerous techniques have been proposed in the formal method community to improve solver efficiency through conflict analysis~\cite{marques2009conflict,heule2011cube}, structural analysis~\cite{wu2007qutesat,zhang2021circuit}, etc. However, lifting these techniques into a privacy-preserving setting remains an open challenge~\cite{luo2022ppsat}.
Although \texttt{ppSAT.Solve} adapts DPLL as its base routine with various branching heuristics for acceleration. As a general-purpose SAT solver, it doesn't exploit structural information derived from hardware designs. Prior work has shown that incorporating structural information into variable ordering can substantially improve solving efficiency in SMT solving~\cite{chen2018control}. To enable \texttt{ppSAT.Solve} to benefit from hardware semantics under encryption, \tool augments the \texttt{ppSAT.Solve} with a hardware
control-flow guided heuristic, referred to as \texttt{hw-ppSAT}.

\subsubsection{Control-flow Guided Heuristic}\label{subsub:heuristic}
Formally, let $\var$ denote the set of all variables in the Boolean formula generated from a hardware IP design. Our heuristic defines a \emph{partial order} relation $\prec$ over $\var$. For any two variables $x, y \in \var$, if $y \prec x$, then $x$ should always be justified before $y$. That is, when both $x$ and $y$ are undecided, and the solver must make a branching decision, it will always assign a value to $x$ before considering $y$. The order captures the three principles as follows:

\noindent \ul{\textit{Principle 1: Variables encoding control signals are prioritized for branching over all other signals.}} In hardware designs, a control signal may govern multiple data paths. Once it is assigned, many downstream data values become either constants or irrelevant to the evaluation, allowing the solver to prune a large portion of the search space.

\begin{figure}[h]
    \centering
        \begin{minipage}{0.9\columnwidth}
            \centering
            \begin{subfigure}[b]{0.45\columnwidth}
                \centering
                \includegraphics[width=\columnwidth]{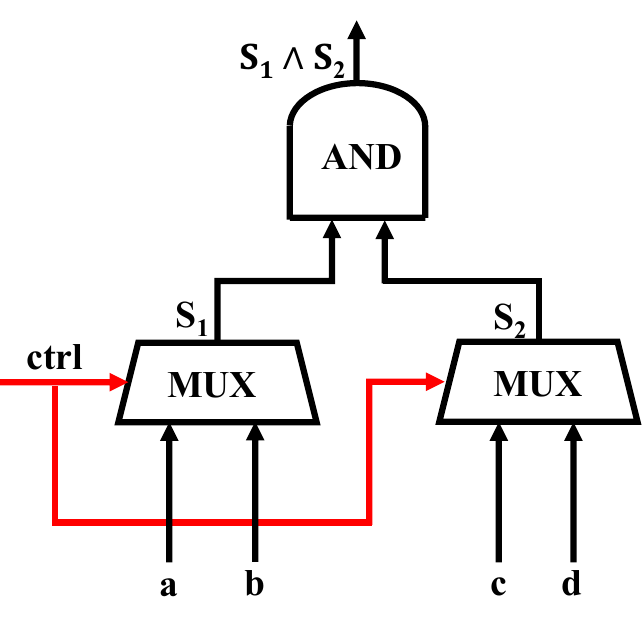}
                \caption{\footnotesize Heuristic 1: \signalname{ctrl} is prioritized}
                \label{fig:subfigheu1}
            \end{subfigure}
            \hfill
            \begin{subfigure}[b]{0.45\columnwidth}
                \centering
                        \includegraphics[width=0.6\columnwidth]{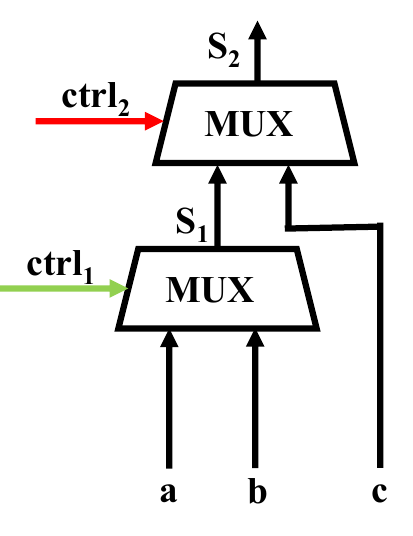}
                \caption{\footnotesize Heuristic 2: \signalname{ctrl2} dominates \signalname{ctrl1}}
                \label{fig:subfigheu2}
            \end{subfigure}
        \end{minipage}
        \caption{Illustration of two signal prioritization principles. In (a), the control signal \signalname{ctrl} is prioritized during branching, since assigning \signalname{ctrl} to \texttt{false} renders the values of \signalname{a} and $\signalname{c}$ irrelevant to the satisfiability of the encoded design. In (b), for two control signals \signalname{ctrl1} and \signalname{ctrl2}, \signalname{ctrl2} is prioritized, because once \signalname{ctrl2} is assigned \texttt{false}, \signalname{ctrl1} no longer influences the evaluation.}
    \label{fig:main}
\end{figure}

\smallskip
\noindent{\it Example.} Figure~\ref{fig:subfigheu1} depicts the gate structure for \(\signalname{S}_1 \wedge \signalname{S}_2\), where the values of \(\signalname{S}_1\) and \(\signalname{S}_2\) depend on the signal \(\signalname{ctrl}\). When \(\signalname{ctrl}\) is true, \(\signalname{S}_1\) and \(\signalname{S}_2\) propagate the values of \(\signalname{a}\) and \(\signalname{c}\), respectively; otherwise, \(\signalname{b}\) and \(\signalname{d}\) are passed. Without prioritizing the control signal \(\signalname{ctrl}\), verifying \(\signalname{S}_1 \wedge \signalname{S}_2\) requires evaluating four possible cases: \(\signalname{a} \wedge \signalname{c}\), \(\signalname{a} \wedge \signalname{d}\), \(\signalname{b} \wedge \signalname{c}\), and \(\signalname{b} \wedge \signalname{d}\). Two of these lead to conflicts. However, when \(\signalname{ctrl}\) is fixed, the conflicting cases are eliminated, leaving only two scenarios to verify.

Formally, let $\var_C \subseteq \var$ be the set of \emph{control signals}, and 
$\var_D = \var \setminus \var_C $ be the set of \emph{non-control signals}.
$
\forall\, c \in \var_C,\; \forall\, d \in \var_D : d \prec c.
$


\noindent \ul{\textit{Principle 2: Within a single time frame, outer control signals dominate inner control signals.}} In a nested structure, an outer control signal governs the scope that encloses the logic of inner signals. Evaluating an inner control before its corresponding outer signal leads to unnecessary computation, because if the outer control disables that block, the inner assignment becomes irrelevant and its effect is masked by the outer logic.

\smallskip
\noindent{\it Example.} If $\signalname{ctrl}_2$ is false in Figure~\ref{fig:subfigheu2}, the value of $\signalname{S}_2$ is determined by $\signalname{c}$, while $\signalname{ctrl}_1$ does not affect the result due to the blocked data flow from $\signalname{a}$ and $\signalname{b}$. Therefore, the interpretation of $\signalname{ctrl}_1$ is dominated by $\signalname{ctrl}_2$.

\smallskip
\noindent \ul{\textit{Principle 3: Across unrolled time frames, control signals at later time stages dominate those at earlier time stages.}} When a design is unrolled, signals at earlier time stages are structurally nested within those of later time stages, making them analogous to inner controls. Following the same prioritisation rationale as Principle 2 and classic tuning practices \cite{shtrichman2000tuning}, the solver should evaluate later-stage control signals before earlier ones. Formally, for  signals $c_i \in \var_C$ at time $i$ and $c_j \in \var_C$ at time $j$, if $j > i$, then $c_i \prec c_j$, meaning that $c_j$ should be interpreted first.

\begin{figure}[h]
    \centering
    \includegraphics[width=0.7\columnwidth]{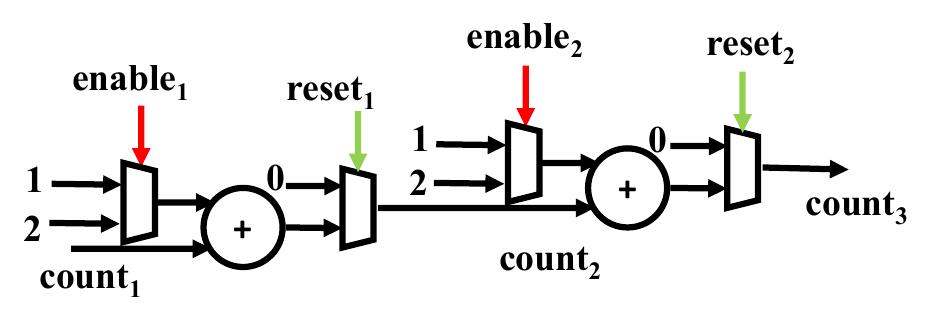}
    \caption{When the circuit is unrolled, later-stage control signals dominate earlier-stage ones. Thus $\signalname{enable}_2$ dominates $\signalname{reset}_1$. Based on  Principle 2, $\signalname{reset}_1$ dominates $\signalname{enable}_1$. Consequently, the priority chain becomes $\signalname{enable}_2 \succ \signalname{reset}_1 \succ \signalname{enable}_1$.}
    \label{fig:subfigheu3}    
\end{figure}

Figure~\ref{fig:subfigheu3} illustrates the unrolled gate structure of the $\signalname{counter}$ signal from Example~\ref{fig:rtlcounter}. In this case, the symbolic value of $\signalname{count}_3$ is updated from $\signalname{count}_1$ through the unrolled circuit. To verify a property involving $\signalname{count}_3$, the solver should justify $\signalname{reset}_2$ and $\signalname{enable}_2$ before $\signalname{reset}_1$ and $\signalname{enable}_1$.

In general, determining the partial order set of control signals from a CNF is non-trivial because structural information disappears once the design is encoded into Boolean clauses. \tool allows the IP vendor to determine the partial-order set locally during preprocessing (see Section~\ref{subsec:ipdesign}), without leaking any sensitive design information to the IP user. 

Specifically, control signals are extracted through AST traversal. The prioritisation of control signals is derived from dataflow analysis using Depth First Search (DFS). For signals $c_1, c_2 \in \var_C$ within a single time frame,
$
c_1 \prec c_2 \; \text{if} \; \depth(c_1) > \depth(c_2),
$
where $\depth: \var_C \rightarrow \mathbb{N}$ maps each signal to its depth obtained by the DFS from dataflow. Both extraction and prioritisation can be performed based on RTL static analysis tools such as ~\cite{takamaeda2015pyverilog,wolf2016yosys}.

\subsubsection{\texttt{hw-ppSAT} Implementation}\label{subsub:hwppsat}

Unlike \texttt{ppSAT.Solve}, \texttt{hw-ppSAT} requires the IP user to provide a partial-order set $\heuset$ as an additional input, which is generated on the IP vendor side.

To prevent information leakage, the heuristic set associated with each IP design is retrieved through the same Oblivious Design Selection protocol described in Section~\ref{subsec:ot_selection}. Concretely, the IP vendor generates a heuristic set $\heuset^k$ for each design $\phi_k$ during preprocessing and includes it in the message vector used in the OT protocol. The IP user then retrieves the corresponding heuristic set using a 1-out-of-$N$ OT via
$$
(\heuset^i, \bot) \leftarrow {\sf OT}(i, \{\heuset^1, \heuset^2, \ldots, \heuset^N\}).
$$
By the security guarantee of OT, the user learns only $\heuset^i$ and obtains no information about $\{\heuset^j\}_{j\ne i}$, while the vendor learns nothing about the user's selection $i$. Furthermore, since $\heuset$ only specifies a partial ordering of variables for branching decisions, it does not reveal structural information about the underlying circuit.

\begin{algorithm}[h]
\caption{Decision Step in \texttt{hw\--ppSAT}}
\label{alg:ctrl}
\begin{algorithmic}[1]
\Require $L = \{\ell_1, \ldots, \ell_{2n}\}$, $C = \{C_1, \ldots, C_m\}$, $\heuset$
\Ensure $d \in [1..n] \times \{0,1\}$ 
\State $ d \gets (\bot, \bot)$, $ d' \gets (\bot, \bot)$
\Comment{$d$: decision; $d'$: fallback decision}

\For{$i \in \heuset $}
    \State $s_P \gets \bot,\; s_N \gets \bot$
    \Comment{Check literal polarity in clauses}
    \For{$j \gets 1 \text{ to } m$}
        \State $s_P \gets s_P \lor C_j.\text{contain}(\ell_i)$
        \State $s_N \gets s_N \lor C_j.\text{contain}(\lnot \ell_i)$
    \EndFor
    \State $d \gets s_N\ ?\ (i, 0) : d$
    \State $d \gets s_P\ ?\ (i, 1) : d$
\EndFor

\For{$i \in  L\setminus \heuset $}
    \Comment{Fallback to standard DLIS heuristic for non-control signals}
    \State $ d' \gets \textsc{DLIS}(C,i)$
\EndFor

\State $d \gets (d \neq (\bot,\bot))\ ?\ d : d'$
\Comment{Use heuristic decision; otherwise, fallback}
\end{algorithmic}
\end{algorithm}
We formally define the \texttt{hw-ppSAT} decision step in Algorithm~\ref{alg:ctrl}, which implements a hybrid method that combines DLIS (the best in \texttt{ppSAT.Solve}) with a control-flow-guided heuristic. In this approach, \texttt{hw-ppSAT} selects a control literal available from \(\heuset\) for the branching heuristic. If all control literals have been evaluated and the formula remains unsolved, the solver defaults to the DLIS strategy for the remaining non-control literal. 


\begin{table}[h]
\centering
\caption{Soft IP metadata in each record.}
\label{tab:softip-attributes}
\begin{tabularx}{.7\columnwidth}{|lXl|}
\toprule
\textbf{Attribute} & \textbf{Description}&\textbf{Size} \\
\midrule
\texttt{ID}     & Unique IP identifier & 32B\\
\texttt{Hash}   & Hash of source code  &  32B\\
\texttt{From}   & Vendor's address & 20B\\
\texttt{To}     & User's address   &  20B\\
\texttt{Reference}  & Integrated IDs (up to 5) &32B $\times$ 5 \\ 
\midrule
\texttt{Total}  &  & 264B \\ 
\bottomrule
\end{tabularx}
\end{table}

\subsection{License Authorization}\label{subsec:ip-traceability}
To ensure traceability and transparency throughout the lifecycle of a soft IP, its record is published to the distributed ledger. This record contains the IP’s ownership information, transaction details, and references to its integrated components.
The format of the records is detailed in Table~\ref{tab:softip-attributes}. 
\input{revision/redistribution}

%% file: revision/redistribution.tex
However, these ledger-based mechanisms do not physically prevent redistribution outside the marketplace. Redistribution risks can be mitigated by integrating complementary protection techniques such as hardware locking~\cite{kahng2001constraint}, active metering~\cite{alkabani2007active}, and watermarking or fingerprinting methods~\cite{chakraborty2009harpoon,kamali2022advances}, where the buyer’s ownership information can serve as a unique identifier embedded into the distributed IP instance.

%% file: sections/optimization.tex
\section{Design Pruning}\label{subsec:prune}
Encoding a design as a pair of binary matrices $(P, N) \in \{0,1\}^{n \times m}$ for property verification can sometimes incur unnecessary cryptographic overhead, as not all variables and clauses derived from the design are relevant to the property.
To reduce the matrices' size, design pruning is applied before oblivious IP selection.
As described in Section~\ref{subsec:ipdesign}, each design statement is translated into an SMT representation. Given a specific property, the IP vendor can perform a Cone-of-Influence reduction~\cite{berezin1997compositional} through backtracking the transitive fan-in from the property signal and retaining only the property-relevant parts of the design to minimize $m$ and $n$. \input{revision/COI}

Moreover, pruning is performed offline on the vendor side based on the user-provided property signal. Although the semantics of the property remain private to the user, the signal itself must be revealed to the vendor for COI analysis. To mitigate this disclosure, users may first submit pruning requests for multiple candidate designs as decoys to hide the actual verification target and then proceed with verification for the selected design.

%% file: revision/COI.tex
SAT-based model checking has proved
that COI reduction is particularly effective for control-dominated designs
where the property depends on a small subset of signals, but provides
limited reduction for datapath-intensive circuits where dependencies rapidly
propagate through the design~\cite{mcmillan2002applying,clarke1997model}. In Section~\ref{subsec:performance}, our experimental results further confirm this observation.

%% file: sections/experiments.tex
\section{Experiment}\label{sec:exp}
\noindent
{\bf Testbed.}
We implement OT and 2PC components (oblivious design selection and secure verification) of the framework using the semi-honest 2PC library of the EMP-toolkit~\cite{wang2016emp}. We build our static analysis frontend on top of PyVerilog~\cite{takamaeda2015pyverilog} and apply Z3~\cite{de2008z3} to generate Boolean formula for both the designs and the properties. The experiment is conducted on an AWS \texttt{z1d.3xlarge} with 12 vCPUs and 96 GB memory.

\noindent
{\bf Benchmarks.}
Table~\ref{tab:source} lists all benchmark descriptions and detailed assertions.
We verify bounded \textit{assert properties} and \textit{cover properties} on ANSI-C~\cite{mtk2016}, OpenCores~\cite{opencores} and Trusthub~\cite{TrustHub}. ANSI includes the corresponding assertions within the code, and Trusthub, where chip-level designs
are annotated with known vulnerabilities that can be expressed as \textit{cover property}.
For example, we verify constraints on output signals and state transitions to ensure correct sequential behaviour in the finite state machines \texttt{b01} and \texttt{b02}. The \texttt{Rrobin} design serves as a round-robin arbiter, where mutual exclusion is enforced by asserting that two acknowledge outputs should never be high simultaneously. The \texttt{adbg} module, a JTAG-compliant TAP controller, includes a password check mechanism, and we verify that no invalid input can bypass this check to set \texttt{passchk = 1}. In the Trusthub benchmark, \texttt{RS232\_T700} verifies whether an input sequence causes the transmitter to enter a Denial-of-Service (DoS) state. In \texttt{memctrl\_T100}, we check whether writing \texttt{0x77} in the CSR register results in the flash sleep bit (\texttt{fs}) being erroneously set to 1. Lastly, \texttt{wb\_conmax\_T300} checks whether any crafted input pattern allows a master to alter its destination slave unexpectedly.

\input{sections/benchmark}
\subsection{Performance Evaluation}\label{subsec:performance}
\begin{figure}[!t]
    \centering
    \includegraphics[width=0.6\linewidth]{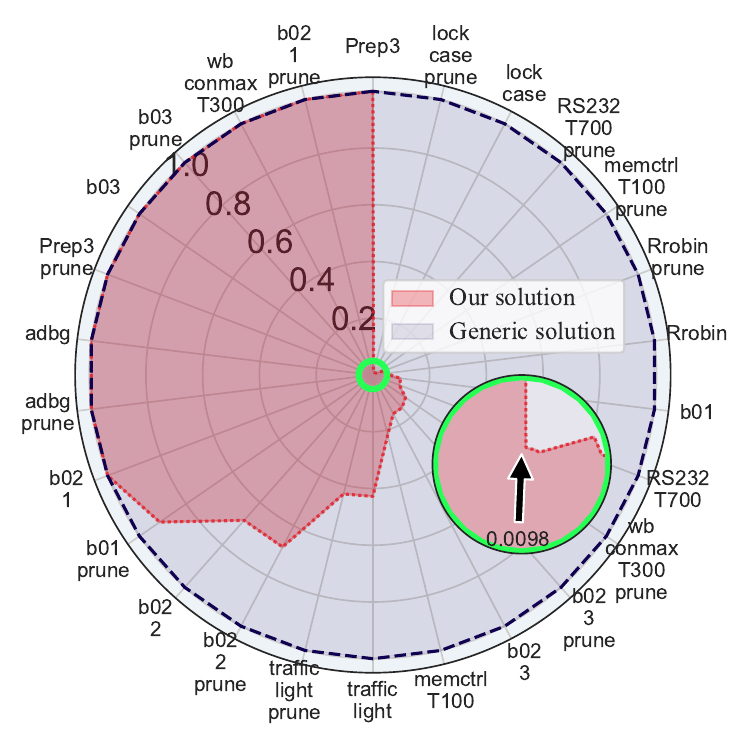}
    \caption{Illustrates the performance ratio between (generic)\texttt{ppSAT} and (our) \texttt{hw-ppSAT} across various (pruned) benchmarks. Each segment represents a specific design, with the outer boundary (ratio = 1.0) indicating the baseline solving time of \texttt{ppSAT}. The shaded region corresponds to \texttt{hw-ppSAT}; a smaller area indicates a greater speedup. As shown, most designs experience a substantial reduction in solving time, with normalized ratios significantly below 1.0. In the best-case scenario, our method achieves a remarkable speedup with a ratio as low as 0.0098.}
    \label{fig:overall}
\end{figure}

\input{sections/ppsat_comparison_table}
We highlight two types of performance improvements in this evaluation: (1) time reductions due to design pruning, and (2) improvements from \texttt{hw-ppSAT}.
For 13 IPs and their corresponding 26 (pruned) design pairs, 7 out of 13 IPs benefit directly from design pruning, achieving noticeable reductions in both variable and clause counts. In addition, 17 out of 26 instances (13 design pairs) demonstrate improved verification time with our control-flow-guided heuristics, as shown in Figure~\ref{fig:overall}.
Compared to the total solving time, the overhead introduced by the IP selection and demasking phases is negligible—typically accounting for less than a few percent of the overall runtime. This indicates that the main performance bottleneck lies in the SAT solving stage, where our heuristics yield the most substantial speedup.
Although the overall verification time can still be long for large and complex IPs, this cost is one-time and occurs only during the initial verification phase before the IP transaction. 


\noindent\textbf{Evaluation of Design Pruning}
We first evaluate the impact of pruning with \texttt{ppSAT}. Table~\ref{tab:exp} also reports the corresponding pruning-induced improvements for \texttt{hw-ppSAT}. Since pruning is solver-agnostic and fundamentally reduces the size of the input matrices processed by the solver, its impact is similar for both solvers, and we therefore omit a separate discussion.

As shown by the highlighted entries in Table~\ref{tab:exp}, \texttt{ppSAT} achieves significantly faster performance on the pruned designs in 8 out of 13 benchmark pairs. This improvement is driven by the reduced formula size, which lowers both the overhead of oblivious design selection (OT time) and the 2PC workload (Solving time).

In particular, benchmarks such as \texttt{b01} and \texttt{Prep3} contain extraneous state machine logic that is not exercised by the property, allowing pruning to reduce the SAT solving time from 51853.60s to 2667.76s in \texttt{b01}, and from 20561.45s to 15176.60s in \texttt{Prep3}.
The large IPs such as \texttt{memctrl\_T100} and \texttt{wb\_conmax\_T300} with multiple control/data paths achieve orders-of-magnitude speedups.
For \texttt{memctrl\_T100}, both cases reach a timeout, while the timeout threshold is reduced from 14 million seconds to 24 thousand seconds.
In the case of \texttt{RS232\_T700}, although the design initially contains 1487 variables, only a subset related to the denial of service vulnerability is relevant, allowing a drastic formula reduction to just 225 variables and a run-time drop from 163885s to 31603.80s.
In contrast, smaller designs such as \texttt{b02\_2}, \texttt{b03}, \texttt{Rrobin}, and \texttt{lock\_case} exhibit little to no performance change, as they are single-purpose IPs with compact state machines that leave no room for structural reduction.




\noindent\textbf{Evaluation of \texttt{hw-ppSAT}}\label{exp:hwppsat}
The integration of control-flow-guided heuristics in \texttt{hw-ppSAT} delivers a clear performance advantage across a majority of benchmarks. As seen in Table~\ref{tab:exp}, 17 out of 26 instances exhibit faster solving times when switching from the default decision heuristic (\texttt{ppSAT}) to the control-aware version (\texttt{hw-ppSAT}). This speedup is further illustrated in Figure~\ref{fig:overall}. This improvement is primarily driven by the reduced number of \texttt{giant steps}.
For instance, in \texttt{b01\_prune}, \texttt{hw-ppSAT} reduces runtime from 2667.76s to 2435.16s by cutting 147 steps, while in \texttt{RS232\_T700\_prune}, the step count drops from 20000(timeout) to 222, resulting in a total runtime reduction from 31603.80s to 350.80s. Similarly, the \texttt{traffic\_light\_prune} benchmark improves from 22330.56s to 9643.22s, and \texttt{memctrl\_T100\_prune} from 24927s to 294.14s. 

In 9 out of 26 instances, no runtime improvement is observed. Comparing the \texttt{Ctrl} and \texttt{DLIS} configurations shows that the proposed heuristic neither improves nor degrades performance in these cases. This behavior is inherent to SAT solving: heuristics are only invoked when the solver must choose the next undecided literal. For these instances, the solver can resolve the formula through propagation without making heuristic-driven decisions, and thus the heuristic is never applied during the solving process.
\subsection{Communication Overhead}\label{subsec:overhead}
\begin{figure}[h]
    \centering
    \begin{subfigure}[t]{0.47\linewidth}
        \centering
       \includegraphics[width=\linewidth]{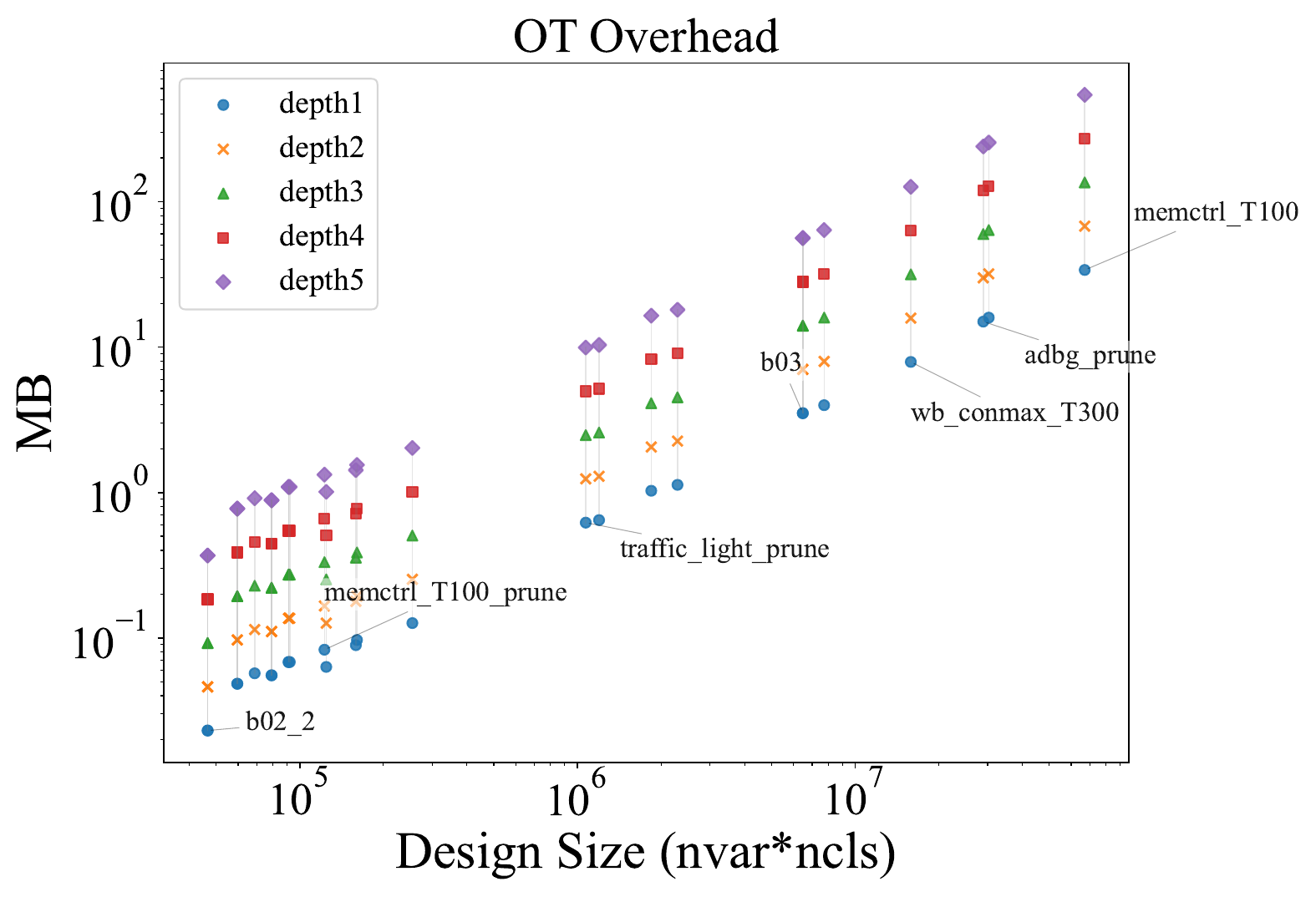}
       \caption{Communication overhead of the $1$-out-of-$2^{\textit{depth}}$ OT protocol versus design size ($n_{var}\times n_{cls}$). The overhead scales approximately linearly with design size and increases with portfolio size $2^{\textit{depth}}$.}
        \label{fig:ot_vs_design}
    \end{subfigure}\hfill
    \begin{subfigure}[t]{0.47\linewidth}
        \centering
        \includegraphics[width=\linewidth]{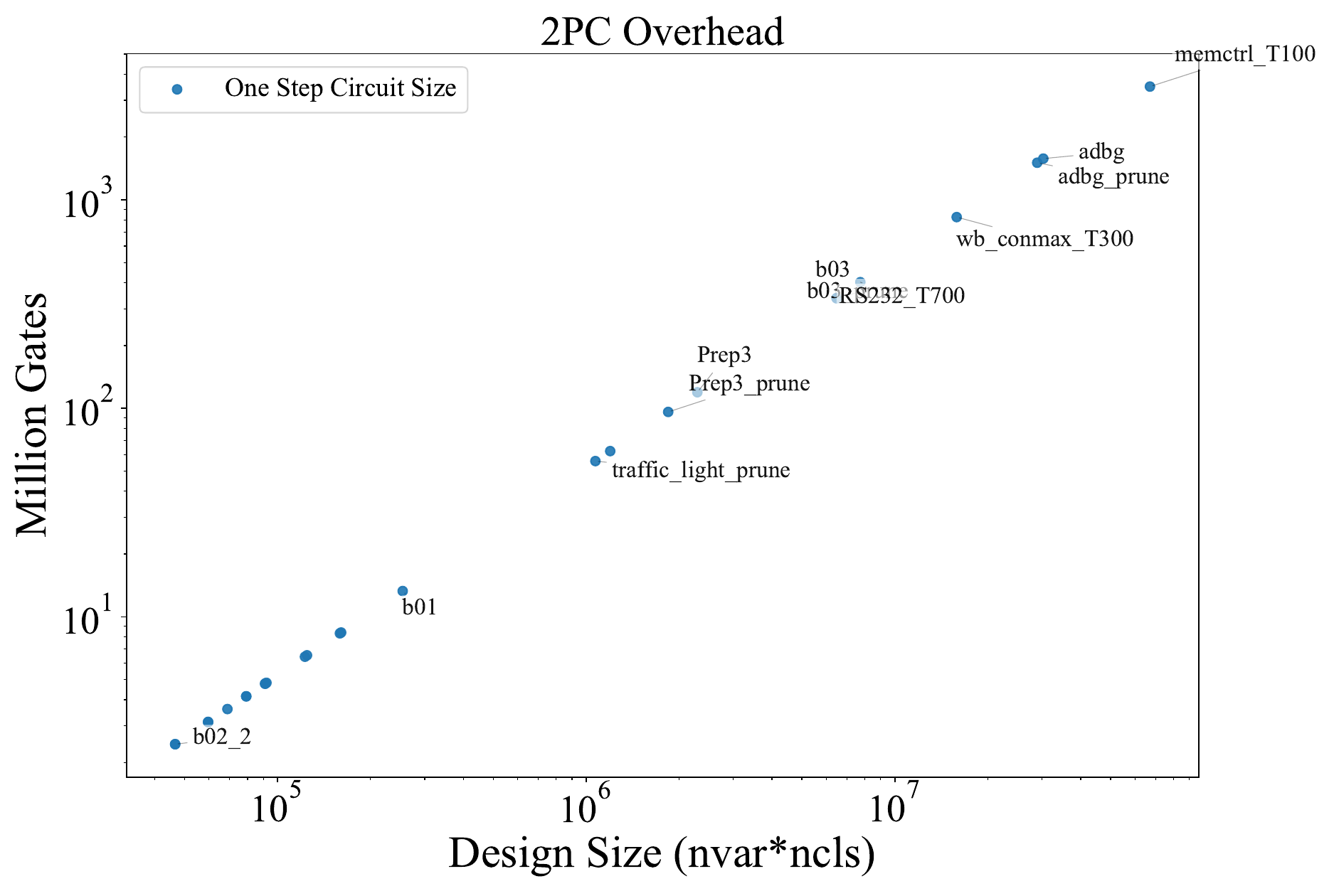}
        \caption{2PC communication overhead per giant step measured by the garbled circuit size. The circuit size grows approximately linearly with the design size ($n_{var}\times n_{cls}$)}
        \label{fig:overhead_all}
    \end{subfigure}
\end{figure}

Figure~\ref{fig:ot_vs_design} illustrates the communication overhead introduced by the $1$-out-of-$2^{\textit{depth}}$ OT protocol. 
The x-axis represents the design size measured as the product of the number of variables and clauses ($n_{var}\times n_{cls}$), while the y-axis reports the total OT cost in MB. Each point corresponds to a verification instance.
As shown in Figure~\ref{fig:ot_vs_design}, the OT communication overhead increases approximately linearly with the design size. As discussed in Section~\ref{subsec:overviewppsat}, each clause in the CNF-encoded design is represented as a pair of binary vectors of length $n$. Consequently, the OT overhead scales with $n_{var} \times n_{cls}$.

Furthermore, the OT overhead grows approximately linearly with $2^{\textit{depth}}$. In a $1$-out-of-$2^{\textit{depth}}$ OT protocol, increasing $\textit{depth}$ enlarges the portfolio size $N = 2^{\textit{depth}}$. As a result, more masked candidate messages must be communicated during the transfer. For the smallest benchmark, \texttt{b02}, with $\textit{depth}=1$, the OT communication overhead is approximately 0.022~MB, whereas for the largest benchmark, \texttt{mem\_ctrl}, with $\textit{depth}=5$, the OT overhead reaches approximately 517~MB.

Figure~\ref{fig:overhead_all} shows the cryptographic overhead of 2PC. This overhead primarily arises from \texttt{hw-ppSAT}, which adapts the DPLL algorithm as its core routine and compiles each \textit{giant step} into the garbled circuit, where the communication overhead scales linearly with the number of AND gates (about 32 bytes per gate). Thus, the total communication cost is the per-step circuit cost multiplied by the number of \textit{giant steps}.
Similar to the OT protocol, the circuit size grows approximately linearly with the design size. Larger designs produce larger CNF encodings and require more evaluation during verification. For the largest design, \texttt{mem\_ctrl}, the circuit size can reach around 3 billion gates per step, making secure verification impractical without optimization. After applying design pruning, we reduce the communication overhead to about 6 million gates per step. Furthermore, using the control-flow-guided heuristic, the number of giant steps is reduced from more than 20,000 to 236, making secure verification feasible.

\subsection{Comparison with Non-private Baseline}\label{subsec:nonprivate}
\begin{figure}[t]
    \centering
    \includegraphics[width=0.6\linewidth]{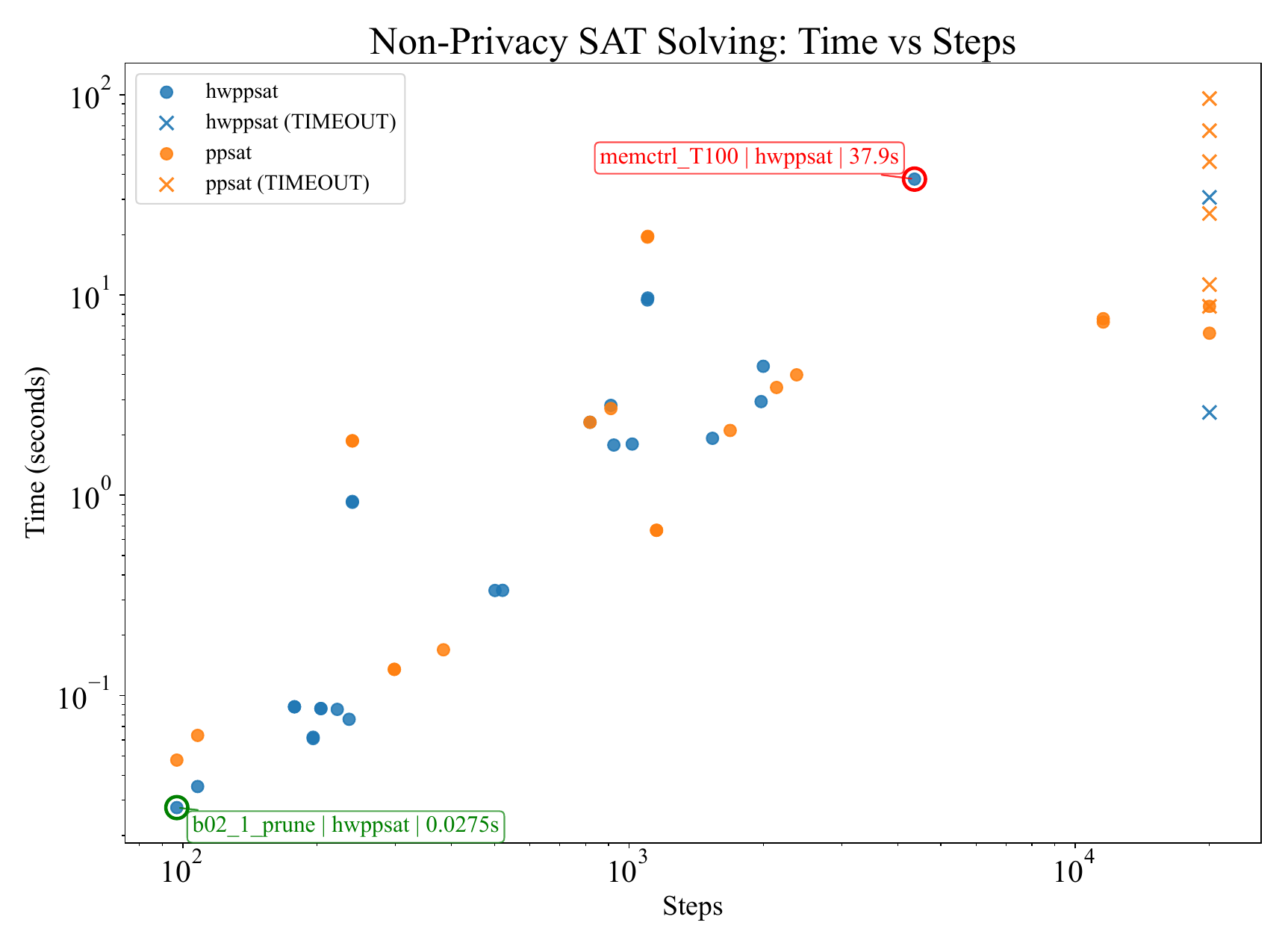}
    \caption{Runtime of the non-private baseline implementation of \texttt{hw-ppSAT} without 2PC. The results indicate that solver runtime is largely dominated by the number of executed steps.}
    \label{fig:nonppSAT}
\end{figure}
We implement \texttt{hw-ppSAT} and \texttt{ppSAT} in Python and include the non-private baseline experiment in Figure~\ref{fig:nonppSAT}. To ensure consistency with Table~\ref{tab:exp}, the timeout is set to 20,000 giant steps, identical to that used in Table~\ref{tab:exp}. Figure~\ref{fig:nonppSAT} shows that reaching 20,000 giant steps corresponds to approximately 100 seconds for the largest timed-out design. The results also indicate that solver runtime is largely dominated by the number of executed giant steps, as the baseline implementations follow the same DPLL core routine.
The most time-consuming benchmark, \texttt{memctrl\_T100}, finishes verification in 37.9 seconds, corresponding to an approximately $8.4 \times 10^{4}$ slowdown compared with the non-private baseline. In contrast, \texttt{b02\_1\_prune} finishes within 0.02 seconds, incurring an overhead of about $3 \times 10^{3}$. As expected, the non-private verification implementation is significantly faster than the privacy-preserving one due to the additional cryptographic operations. Moreover, the slowdown is not uniform across benchmarks and primarily depends on the design size, as larger designs incur substantially higher overhead in GC-based 2PC computation.

%% file: sections/benchmark.tex
\newcolumntype{Y}{>{\raggedright\arraybackslash}X} %
\newcolumntype{T}{>{\ttfamily\raggedright\arraybackslash}X} %
\begin{table*}[h]
\centering
\caption{Design, Line of Code, Abstract Description and Assertion. We expect \textsc{UNSAT} for the negation of the assertion property conjoined with the design, and \textsc{SAT} for the counterexample derived from the conjunction of the cover property and the design.
}
\scriptsize
\begin{tabularx}{\textwidth}{l r Y T}
\toprule
\textbf{Benchmark[Source]} & \textbf{LoC} & \textbf{Description} &\textbf{SVA Assertion}\\
\midrule
\texttt{Rrobin}~\cite{mtk2016} & 59 & round robin arbiter& assert (ack0==0 ||  ack1==0)\\
\texttt{b01}~\cite{mtk2016} &  111& 8-state Mealy FSM& assert  (OVERFLW==1 |-> \#\#2 (OVERFLW==0)) \\
\texttt{b02\_1}~\cite{mtk2016} & 81& 7-state Moore FSM &assert (stato==D |-> \#\#2 (stato==B))\\
\texttt{b02\_2}~\cite{mtk2016} & 81& 7-state Moore FSM & assert (U==1 |-> \#\#1 U==0)\\
\texttt{b02\_3}~\cite{mtk2016} & 81& 7-state Moore FSM &assert (U==1 |-> stato==B)\\
\texttt{b03}~\cite{mtk2016} &119 & 4-request arbiter& assert (GRANT\_O==0 || GRANT\_O==8 || GRANT\_O==4 \\
 & & &|| GRANT\_O==2 || GRANT\_O==1)\\

\texttt{traffic\_light}~\cite{mtk2016} & 52 & 3-state FSM with counter& cover (reset\#\#5 (Light\_Sign == YELLOW\_LIGHT) )\\
\texttt{lock\_case}~\cite{opencores} & 25& 4-state Moore FSM &cover (unlock == 1)\\
\texttt{Prep3}~\cite{opencores} & 122 &8-state Mealy FSM&assert  (next\_out == 8’h20 |-> \#\#1 (next\_out == 8’h11\\
&  &&|| next\_out == 8’h30 ))\\
\texttt{adbg}~\cite{opencores} & 531 &  JTAG  controller&cover (next\_passchk == 1) \\
\texttt{memctrl\_T100}~\cite{TrustHub} & 1786 & memory controller&cover (csr\_r[2]==0 \&\& fs ==1)\\
\texttt{RS232\_T700}~\cite{TrustHub} & 231 & uart transmitter&cover (state\_DataSend==7)\\
\texttt{wb\_conmax\_T300}~\cite{TrustHub} &7626  &  general round robin arbiter&cover (slv\_sel != wb\_addr\_i[31:28])\\
\bottomrule
\end{tabularx}
\label{tab:source}
\end{table*}

%% file: sections/ppsat_comparison_table.tex
\begin{table*}[h]
   \scriptsize 
    \begin{center}
    \caption{We summarize the end-to-end execution time of \tool for 13 IP designs in both their original and pruned forms.
\texttt{Var/Cls} denotes the number of variables and clauses in the verification formula $\psi$.
\texttt{Design Size} represents the size of the encoded design transferred during OT, measured in 128-bit blocks.
\texttt{OT(1/32)} reports the execution time of 1-out-of-32 OT.
\texttt{Demask} measures the time required to reconstruct the selected design within the 2PC (Equation~\ref{eq:deobfuscate}).
For the SAT solving stage, \texttt{DLIS} records the number of giant steps taken under the standard \texttt{ppSAT} DLIS heuristic, and \texttt{ppSAT} reports the corresponding solving time.
Similarly, \texttt{Ctrl} denotes the number of giant steps taken under \texttt{hw-ppSAT}, while \texttt{hw-ppSAT} reports the solving time using the control-flow-guided heuristic. The results show that combining design pruning with \texttt{hw-ppSAT} achieves faster verification times on 12 of 13 benchmarks. In the table, \textcolor{yellow}{yellow cells} represent time improvements from design pruning, and \textcolor{green}{green cells} represent time improvements from heuristic guidance.} \label{tab:exp}
\resizebox{\textwidth}{!}{%
\begin{tabular}{p{2.3cm}p{.5cm}p{.5cm}p{.5cm}lllllll}
\toprule
              Design & Bound &  Var &   Cls & Design Size(blocks) & OT (1/32)(ms) & Demask(s) &  DLIS  &     ppSAT(s) & Ctrl &   hw-ppSAT(s) \\
\midrule
               Prep3 &     3 &  876 &  2617 &    35266 &            146.37 &              80.06 &               911 &  20561.45 &        911 &   20561.45 \\
         Prep3\_prune &     3 &  771 &  2392 &    32116 &            142.23 &              63.56 &                818 &  \cellcolor{yellow!30}15176.60 &        818 &  15176.60 \\
          RS232\_T700 &     5 & 1487 &  5200 &   124512 &            513.56 &             290.69 &            20000$^*$ &   1638850 &       2000 &     \cellcolor{green!25}163885 \\
    RS232\_T700\_prune &     5 &  225 &   708 &     2784 &             12.33 &               5.99 &            20000$^*$ &  \cellcolor{yellow!30}31603.80 &        222 &   \cellcolor{green!25}  350.80 \\
              Rrobin &     6 &  154 &   387 &     1512 &              4.97 &               2.03 &              11561 &   6704.32 &        521 &   \cellcolor{green!25}  302.13 \\
        Rrobin\_prune &     6 &  154 &   387 &     1512 &              4.25 &               2.04 &             11561 &  6675.69 &        501 &    \cellcolor{green!25} 289.29 \\
                adbg &     4 & 2186 & 13844 &   498168 &           2115.35 &            1148.19 &              1100 & 356573.80 &       1100 &  356573.80 \\
          adbg\_prune &     4 & 2103 & 13756 &   467500 &           1877.73 &            1094.98 &          1102 & \cellcolor{yellow!30}346138.20 &       1102 &  346138.20 \\
                 b01 &     4 &  256 &   994 &     3952 &             16.81 &               9.52 &               20000$^*$ &  51853.60 &       1978 &   \cellcolor{green!25} 5128.32 \\
           b01\_prune &     4 &  211 &   762 &     3024 &             12.08 &               6.28 &               1686 & \cellcolor{yellow!30}  2667.76 &       1539 &   \cellcolor{green!25} 2435.16 \\
               b02\_1 &     3 &  163 &   565 &     2132 &              9.28 &               3.33 &               108 &     97.80 &        108 &      97.80 \\
         b02\_1\_prune &     3 &  144 &   478 &     1784 &              9.41 &               2.47 &              97 &   \cellcolor{yellow!30}  65.73 &         97 &      65.73 \\
               b02\_2 &     2 &  118 &   395 &      720 &              3.80 &               1.64 &              298 &    138.30 &        204 &     \cellcolor{green!25} 94.67 \\
         b02\_2\_prune &     2 &  118 &   395 &      720 &              3.17 &               1.62 &             298 &    140.27 &        204 &    \cellcolor{green!25} 96.02 \\
               b02\_3 &     3 &  160 &   570 &     2132 &              8.77 &               3.24 &             1153 &   1080.95 &        178 &    \cellcolor{green!25} 166.88 \\
         b02\_3\_prune &     3 &  160 &   570 &     2132 &              8.37 &               3.39 &            1153 &   1055.94 &        178 &    \cellcolor{green!25} 163.02 \\
                 b03 &     7 & 1289 &  5026 &   109670 &            436.25 &             223.30 &            240 &  15413.88 &        240 &   15413.88 \\
           b03\_prune &     7 & 1289 &  5026 &   109670 &            430.14 &             226.00 &              240 &  15234.12 &        240 &   15234.12 \\
           lock\_case &     4 &  176 &   450 &     1728 &              9.43 &               2.87 &             20000$^*$ &  15844.80 &        196 &    \cellcolor{green!25} 155.28 \\
     lock\_case\_prune &     4 &  176 &   450 &     1728 &              7.36 &               2.92 &            20000$^*$ &  15661.34 &        196 &   \cellcolor{green!25}  153.48 \\
        memctrl\_T100 &     1 & 5183 & 12943 &  1059194 &           4469.61 &            2450.10 &          20000$^*$ &  14628460 &       4365 & \cellcolor{green!25}3192661.40 \\
  memctrl\_T100\_prune &     1 &  268 &   458 &     2592 &             10.66 &               4.44 &           20000$^*$ &    \cellcolor{yellow!30} 24927 &        236 &    \cellcolor{green!25} 294.14 \\
       traffic\_light &     5 &  590 &  2028 &    20190 &             91.33 &              41.60 &            2376 &  28052.48 &       1017 &  \cellcolor{green!25} 12007.31 \\
 traffic\_light\_prune &     5 &  550 &  1948 &    19390 &             77.22 &              36.82 &              2142 &  \cellcolor{yellow!30}22330.56 &        925 &   \cellcolor{green!25} 9643.22 \\
      wb\_conmax\_* &     1 & 2781 &  5705 &   246972 &           1045.03 &             582.29 &           20000$^*$ &   3264500 &      20000$^*$ &    3264500 \\
wb\_conmax\_*\_prune &     1 &  296 &   421 &     1974 &              8.24 &               3.63 &              2745 &  \cellcolor{yellow!30} 3391.09 &        384 &    \cellcolor{green!25} 474.38 \\
\bottomrule
\end{tabular}
}
  \footnotesize
    \item[$*$]  \textbf{timeout when reaching 20000 giant steps};
    \end{center}

\end{table*}

%% file: sections/conclusion.tex
\section{Conclusion \& Discussion}\label{sec:con}

This paper introduces \tool, the first end-to-end framework for addressing privacy concerns in hardware formal verification and ownership traceability after IP trading. We propose a zero-trust model that eliminates the need for a trusted third party while enabling formal logic verification, a capability not addressed by prior industrial or academic work. By leveraging 2PC and OT, we extend the state-of-the-art privacy-preserving SAT solver \texttt{ppSAT} to the hardware domain (\texttt{hw-ppSAT}) and apply optimizations that make IP-level verification practical. A practical challenge of \tool lies in its interactive protocol, where stages such as oblivious design selection and secure SAT solving require synchronized message exchanges between the IP vendor and the IP user, introducing additional latency compared with offline verification. Despite this overhead, the framework remains compatible with existing post-licensing IP protection solutions and can be integrated into current hardware supply-chain ecosystems.

%% file: sections/ack.tex
\section{acknowledgment}
This work was supported in part by the National Science Foundation under Grant No.~NSF-2304533.